 \pdfoutput=1
\documentclass[pdflatex,sn-mathphys-num]{sn-jnl}

\usepackage{graphicx}%
\usepackage{multirow}%
\usepackage{amsmath,amssymb,amsfonts}%
\usepackage{amsthm}%
\usepackage{mathrsfs}%
\usepackage[title]{appendix}%
\usepackage{xcolor}%
\usepackage{textcomp}%
\usepackage{manyfoot}%
\usepackage{booktabs}%
\usepackage{physics}
\usepackage{algorithm}%
\usepackage{algorithmicx}%
\usepackage{algpseudocode}%
\usepackage{listings}%
\usepackage{natbib}%
\usepackage[T1]{fontenc}

\theoremstyle{thmstyleone}%
\theoremstyle{thmstyletwo}%

\theoremstyle{thmstylethree}%

\begin{document}

\title[Article Title]{Collective emission of subwavelengths atom-like emitter arrays in the presence of inhomogeneous broadening}


\author[1]{\fnm{Uri} \sur{Israeli}}
\equalcont{These authors contributed equally to this work.}

\author[1]{\fnm{Shahar} \sur{Levi}}
\equalcont{These authors contributed equally to this work.}

\author[1]{\fnm{Sagi} \sur{Ben-Avi}}
\equalcont{These authors contributed equally to this work.}
\author[1]{\fnm{Ada} \sur{Kransnovsky}}

\author[1]{\fnm{Daniel} \sur{Silvian}}
\author[1]{\fnm{Shlomo} \sur{Winberg}}
\author[1]{\fnm{Rivka}\sur{Bekenstein}*}

\affil[1]{\orgdiv{Racah Institute of Physics}, \orgname{Hebrew University of Jerusalem}, \city{Jerusalem}, \postcode{91904}, \country{Israel}}


\abstract{
Quantum metasurfaces comprised of subwavelength atomic arrays emerged as a promising platform for enhanced atom-photon interaction. However, realizing such a system with solid-state emitters has been considered impractical due to strong inhomogeneous broadening, which was expected to suppress the photon-mediated interactions that underpin collective emission. Here we report the observation of collective emission from subwavelength arrays of silicon-vacancy centres in diamond --- solid-state emitters whose inhomogeneous broadening exceeds the natural linewidth by two orders of magnitude --- demonstrating that collective effects such as resonance shifts, modified decay rates and directional coherent emission survive this disorder. A crucial enabling element is the implantation of a high density of silicon ions at each array site. This creates so-called superatoms, local ensembles that probabilistically achieve frequency matching across the array and enhance the collective response. We support our observations with a theoretical analysis explaining the mechanisms that preserve the collective effects even in the presence of inhomogeneity. These observations have direct implications for the realization of subwavelength arrays in any solid-state system, paving the way for quantum-emitter metasurfaces that are naturally integrated into nanophotonic environments.
}




\maketitle

\section{Introduction}\label{sec1}

That collective atomic excitation can generate coherent emission of photons with controlled properties is well established \cite{dicke1954coherence, cirac1997Quantum}. In recent years, subwavelength  emitter arrays \cite{asenjo2017exponential, bettles2016enhanced, Shahmoon2017Cooperative, asenjo2019optical} have emerged as a versatile platform for realizing strong atom-photon interaction \cite{bekenstein2020quantum, Rui2020Subradiant,masson2022universality}. The underlying physics of these systems describes the emergence of collective resonances due to exchange interactions between the emitters, mediated by virtual photons. When emitters are placed at subwavelength distances (relative to the wavelength of the exciting light), the exchange interaction of photons between them becomes dominant, and a collective response to light emerges. These interactions are captured by the dyadic Green’s function \cite{Novotny2012} and give rise to frequency shifts of the collective resonances and modified lifetimes relative to the bare emitter resonance. Recent work on the collective emission of subwavelength atom arrays has suggested that the effect is universal in ordered arrays and lattices \cite{sierra2022dicke, masson2022universality}. Such coherent emission has applications in a wide range of scenarios, from the exponential improvement of memory time \cite{asenjo2017exponential} to optical wave-guiding \cite{asenjo2019optical} and entanglement generation between localized qubits \cite{Guimond2019Subradiant, antman2024atom, shah2024quantum}. 

In 2020, the concept of quantum metasurface was proposed \cite{bekenstein2020quantum}, based on the introduction of a single ancillary control atom and the Rydberg blockade mechanism. This theoretical work ignited experimental efforts to realize quantum metasurfaces with ultracold atoms \cite{Rui2020Subradiant}, leading to the implementation of a switchable atomic mirror with arrays of rubidium atoms, in which coherence between a Rydberg excitation and the response of the array to light has been demonstrated \cite{srakaew2023subwavelength}. The prediction of superradiance bursts were parallel observed with atoms interacting in the vicinity of a tapered fibre waveguide \cite{liedl2024observation}. 
In these experiments and other works based on atomic arrays realized with ultracold atoms, the atoms are homogeneous relative to the bare emitter frequency. While efforts exist to integrate ultracold array with nanophotonics \cite{samutpraphoot2020strong,menon2024integrated, liedl2024observation}, solid-state systems, which can be mass fabricated, offer a more natural route to scalability
\cite{guo2024direct,ding2024high}.

However, solid-state systems exhibit natural inhomogeneous broadening that suppresses any photon-mediated interaction that requires frequency-matching. This limitation has been overcome for two-emitter systems by applying strain or by photon frequency shifting \cite{machielse2019quantum, levonian2022optical, evans2018photon, knaut2024entanglement}, but these techniques are impractical for hundreds of emitters, leading to the conception that for solid-state subwavelength quantum emitters featuring strong inhomogeneous broadening collective-emission effects would be suppressed \cite{zhou2025efficient}. 

Here, we report experiments challenging this prevailing view. We observed collective emission from atom-like emitters that have inhomogeneous broadening of two orders of magnitude relative to the bare-emitter decay rate ($\sim$ 10\,GHz), along with collective resonance shifts and decay rates associated with the collective states, which we show to directly determine the emission directionality. The key to this advance is the careful preparation of superatoms --- each array site implanted with a high density of silicon ions --- giving rise to an enhanced collective response and a build-up of coherence of the array excitation even in the presence of strong inhomogeneous broadening. These results establish solid-state emitter arrays as a viable platform for quantum-emitter metasurfaces, opening a route to scalable coherent light sources with designer spectral and directional properties.

\section{Results}\label{sec2}

Our realization employs arrays of silicon vacancy centres (SiVs), which emerged as a promising defect for quantum information. Recent work demonstrated high-fidelity spin–photon gates and entanglement distribution with SiVs \cite{stas2022robust,knaut2024entanglement}, along with long-lived nuclear-spin states practical for quantum memory \cite{Bhaskar2020experimental,stas2022robust}. These advances place the SiV at the forefront for quantum-network applications. 

The SiV centres exhibit four dipole-allowed optical transitions due to the two degenerate doublets that are partially lifted by the spin--orbit interaction (see Fig. 1) \cite{Rogers2026electronic, Goss1996the}.  These four optical transitions have a zero-phonon line with an energy of 1.68 eV (737 nm) and a natural  linewidth of $\sim$94 MHz.

\begin{figure}[h]
\centering
\includegraphics[width=0.9\textwidth]{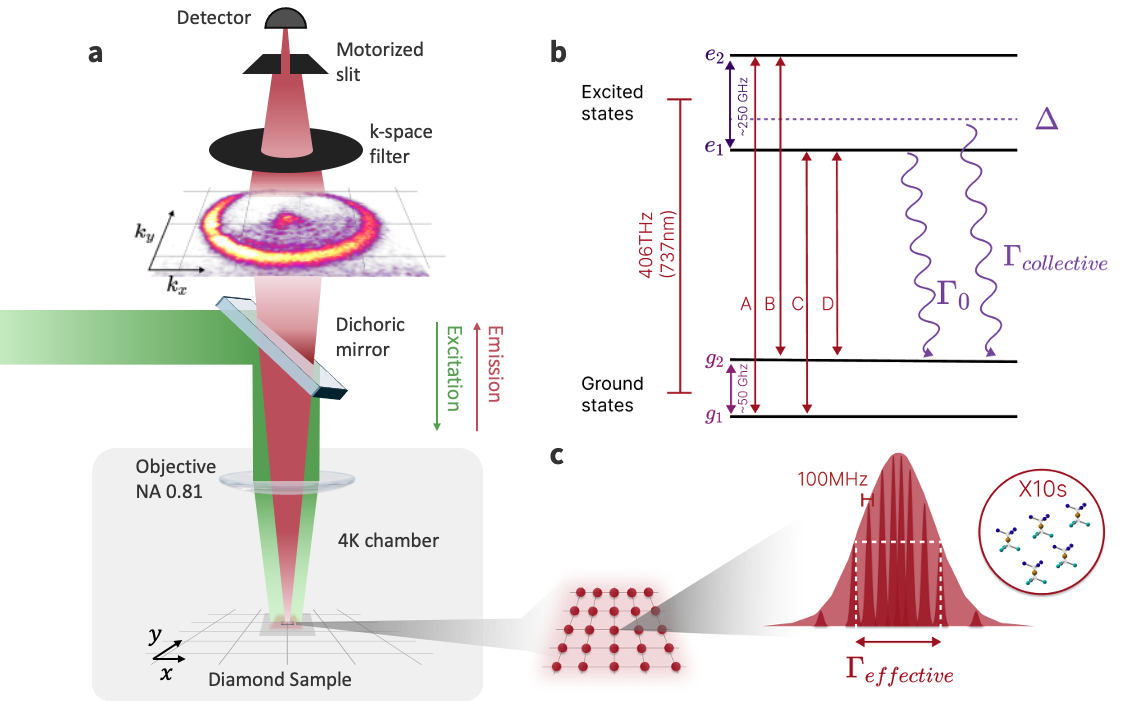}
\caption{\textbf{a}. Experimental setup for probing the collective emission of atom-like emitter arrays. \textbf{b}. The SiV centre is an atom-like emitter system whose ground and excited states are localized at the diamond bandgap. Its spin and optical degeneracies are partially lifted by the spin–orbit interaction, which splits each quartet into two degenerate doublets, so that there are four dipole-allowed optical transitions. These are resonant with light around 737 nm (marked A,B,C,D). We are interested in probing the collective resonance with decay rate $\Gamma_{collective}$ and spectral shift $\Delta$ relative to the bare SiV resonance. \textbf{c}. Each array site is comprised of tens of SiV centres with spectral inhomogeneity that determines the effective decay rate $\Gamma_{eff}$.}\label{fig1}
\end{figure}

To achieve detectable collective emission we prepare subwavelength two-dimensional arrays of silicon-vacancy centres located just below the diamond upper surface. The arrays feature subwavelength spacing that is expected to enhance the collective response --- about 250 nanometers --- which corresponds to about 0.7$\lambda$ ($\lambda$ being the resonant wavelength inside the diamond). To explore the effect of symmetry on the collective emission, we prepare arrays with two distinct symmetries: square and honeycomb. Each array site can be considered a point source as its location is pre-determined by our fabrication process and is defined with 30 nanometer precision.

An important feature of our system has to do with the inhomogeneous broadening of the emitters. Whereas the natural linewidth of the bare SiV is around 100\,MHz, the  natural inhomogeneity of SiVs ensembles is tens of GHz \cite{Evans2016narrow, Aend2026Photoluminescence}. As we are interested in dipole--dipole exchange interaction between emitters, the effects are highly sensitive to frequency matching. To overcome this obstacle and preserve the collective effects we over-implanted silicon ions in each site (see Fig. 1d). This creates an effective superatom that is comprised of many SiVs located effectively at the same point (30\,nm $<<$ wavelength). These SiVs are naturally detuned relative to one another. The superatom is expected to have a broader effective linewidth than that of an individual SiV, due to the natural inhomogeneity of the SiVs. By over-implanting we achieve a double purpose: probabilistic frequency-matching between SiVs in all the array sites (see section 2 in Supplementary Information), and enhancement of the collective response of the array, as detailed in the section on the theoretical model below.

In our experiment, we study the emitter array collective states by probing its directional emission. The coherent emission is a result of correlations that are building up between the emitters. We are interested in observing the collective states of the arrays, which are predicted to be frequency-shifted by $\Delta$ from the bare four optical transition: A,B,C,D (as schematically displayed in Fig. 1). In the experiment the SiV array is cooled to 4K and excited to a high-energy level with a 532\,nm green laser, before it relaxes to one of the doublet excited states. It then decays through a dipole-allowed transition to the doublet ground state \cite{Rogers2026electronic}, after which we observe the emission from the four optical transitions and record their spectrum, focusing our detection on directed emission normal to the array.

\begin{figure}[h]
\centering
\includegraphics[width=0.9\textwidth]{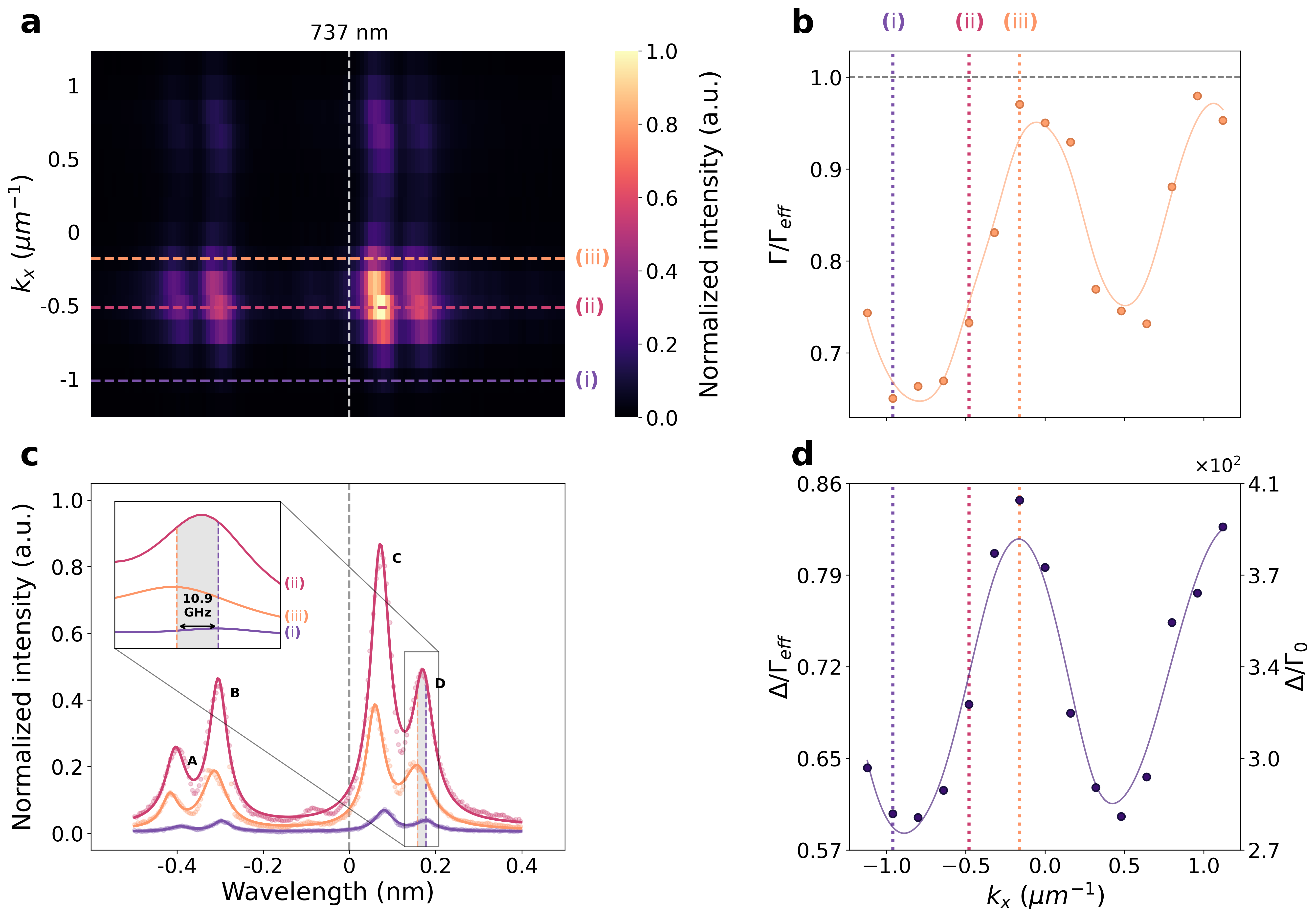}
\caption{\textbf{a}. Spectrum of the collective emission of a SiV subwavelength array, with the four SiV optical transitions as detected in the direction normal to the arrays as a function of the momentum parallel to the array. \textbf{b,d}. The collective linewidth and shifts of the specific collective states emitting as a function of momentum, extracted from the fit for the D transition. \textbf{c}. Cross sections of the spectrum for different momenta, showing the shift in emission frequency, marked (i),(ii),(iii) in all subplots (Inset highlights frequency shift).} \label{fig2}
\end{figure}

\begin{figure}[h]
\centering
\includegraphics[width=0.9\textwidth]{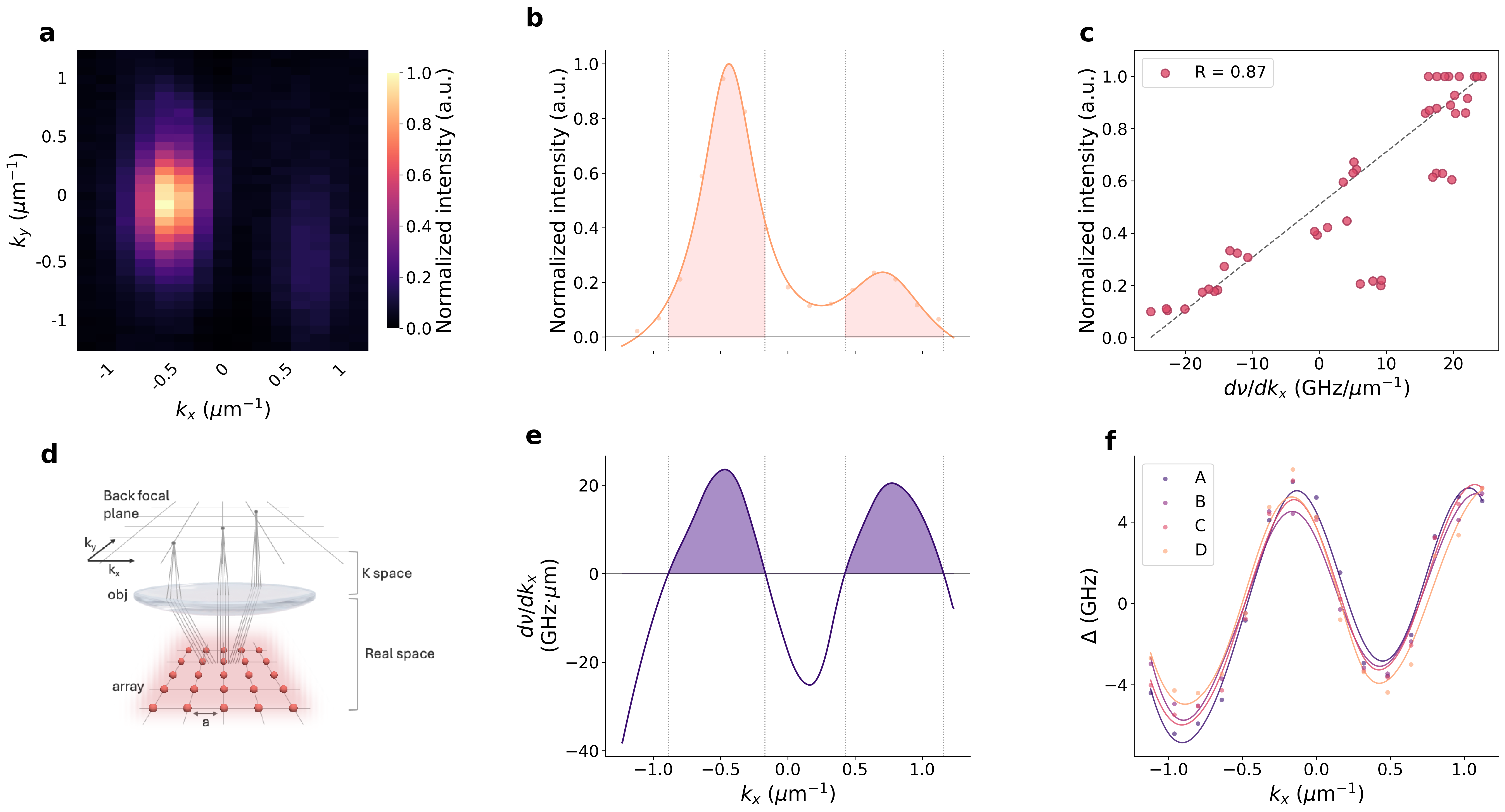}
\caption{\textbf{Directional emission of the collective response of the array.} \textbf{a}. Intensity distribution detected in the direction normal to the array as a function of the momentum parallel to the array. \textbf{b}. Cross sections of the intensity as a function of $k_x$ for different momenta, showing the shift in emission frequency. Red-shaded parts mark high-intensity areas. \textbf{c}. Correlations between the extracted group velocity in \textbf{e} and the intensity in \textbf{b}. \textbf{d}. Schematics of the measurement setup, where a lens maps the emission into momentum space and imaging of the lens back focal plane reveals the intensity distribution in k-space. \textbf{e}. Extracted group velocity in the direction parallel to the array (slope of the band in \textbf{f}). \textbf{f}. Collective resonance shifts of the four SiV optical transitions.} \label{fig3}
\end{figure}

We hereby describe the characteristic of the collective emission which involves many array sites. 
By measuring the spectrum of the emission in momentum space, we find a deviation of the collective resonance frequency from the bare emitters resonance. We record the spectrum as a function of the momentum in the direction parallel to the array. This exposes the photonic band $\omega(k_{\perp})$ for a photon that exists within the array plane, where $\Delta(k_x)$ is defined as the difference from the central resonance peak. When imaging this plane in Fourier space we achieve access to the band information. This band had been previously calculated for specific symmetries (square and one-dimensional array), in our experiment we observe it directly, for two symmetries including the honeycomb that was not previously analyzed. The spectral shift observation reveals the same trend and magnitude for all of the four SiV's optical transitions (see Figure \ref{fig2}a and \ref{fig3}f). %

This result qualitatively resembles the dispersion relation $\omega(k)$ of a one-dimensional chain of emitters that is parabolic near $k_{\perp}=0$ and flattens near the end of the Brillouin zone \cite{asenjo2019optical} (see section 3 of the Supplementary Information). 
The magnitude of the resonance shifts is in the order of the superatom effective linewidth (see Fig. \ref{fig2}d), where in the theoretical prediction it should be in the order of magnitude of the bare emitter linewidth. This strengthens the assumption of the superatom with effective linewidth in each site.  Our experimental scenario is richer than the theoretical analysis, as the SiV has four transitions, and dipole-dipole exchange interaction is not exactly the same for each due to the polarization orientation and electron's orbital shape in the ground state. For the B and D transitions that decay to the second ground state ($\ket{g_2}$) we observe a different linewidth trend than for the A and C transitions that decay to the first ground state ($\ket{g_1}$) (see section 2.1 in the Supplementary).\\

Next, we measure the exact emission directionality. Our setup is designed such that we collect only the emission normal to the array by filtering-out optical modes with high transverse momentum (see Figs \ref{fig1}a and \ref{fig3}d). We image the emitted light's intensity distribution as a function of the transverse momentum in the $k_xk_ y$ plane (see Fig. \ref{fig3}a). As the photons are coherently emitted from the collective array state, involving SiVs from many sites, we find a direct relation between the intensity of the emission and the group velocity determined by the photonic band. Specifically, by calculating the derivative of the frequency as a function of the transverse momentum $\partial\Delta/\partial k_x$ we extract the transverse component of the group velocity (Fig. \ref{fig3}c,d). We then examine the correlation between the latter extracted group velocity and the intensity distribution of the collective emission. Where the group velocity parallel to the array is greater we find higher photon counts. A graph showing the latter correlation is displayed in Fig. \ref{fig3}c, highlighting how an increase in the group velocity leads to larger photon counts in the emission. This correlation demonstrates that our emitter array acts as a photonic metamaterial capable of coherent emission with a spectrum determined by the collective states. This demonstrates how the symmetry of the array and how the characteristics of the emitters directly affect the collective emission direction. The results described up until now establish the possibility of coherent metamaterials as a quantum source with controlled spectrum and directional emission.  \\

\begin{figure}[h]
\centering
\includegraphics[width=0.9\textwidth]{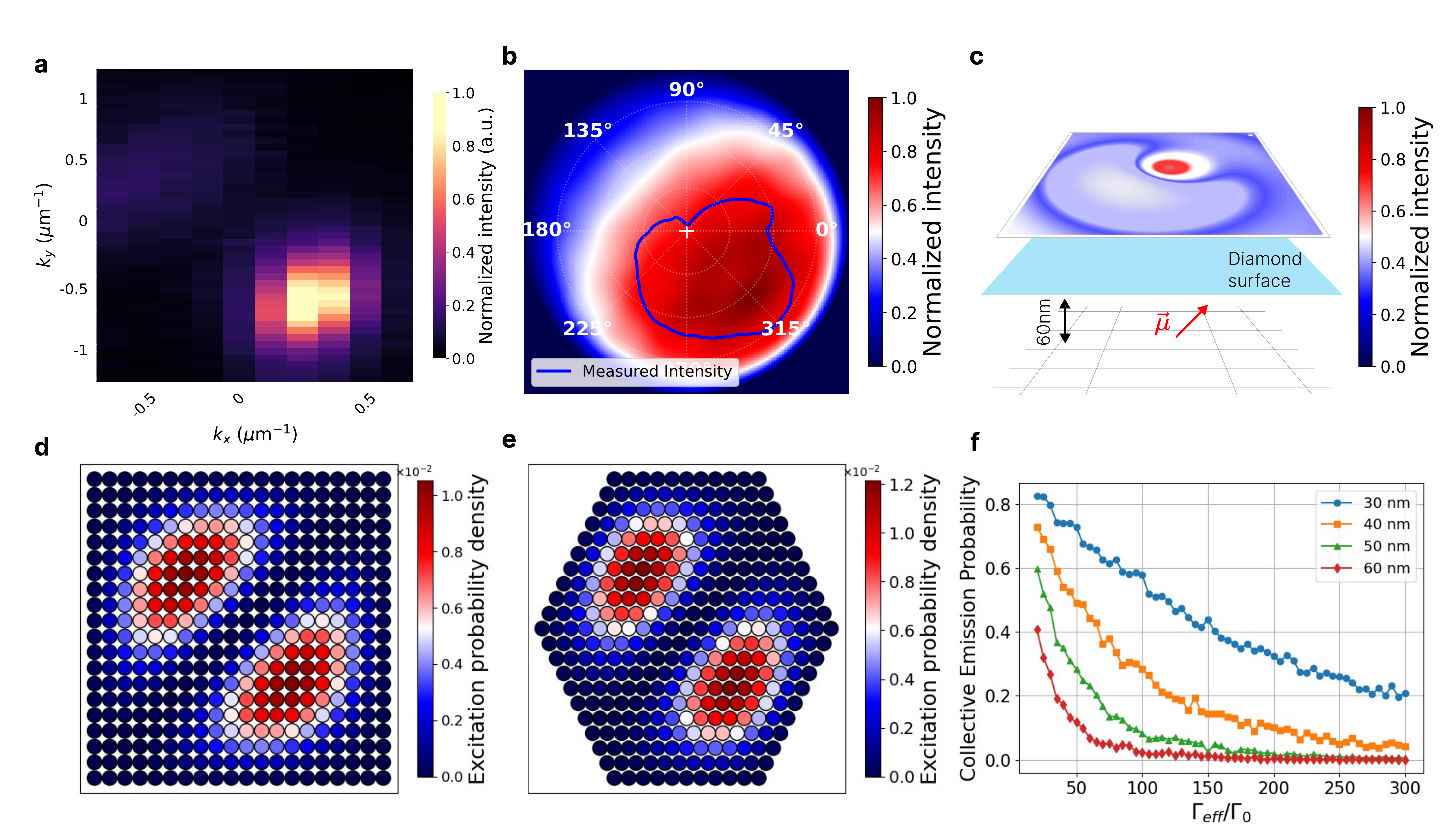}
\caption{\textbf{a}. Measured spatial distribution of the asymmetric collective mode for a square array in the far-field. \textbf{b}. Anisotropic far-field emission of a dipole near the diamond--air interface as calculated by FDTD and compared to the measured uncoupled dipole emission (blue line). \textbf{c}. Schematics of a dipole orientated in the (1,1,1) direction and placed 60\,nm beneath the diamond--air surface (displayed is the calculated electric field perpendicular to the array plane). \textbf{d,e}. Calculated asymmetric subradiant modes of square and honeycomb arrays. \textbf{f}. Probability of superradiance for a superatom of 40 emitters at a location with high emitter density. 20 percent of the superatoms are expected to be superradiant after optical pumping.} \label{fig4} \label{fig4}
\end{figure}

We now focus on the unique spatial distribution of the emission intensity detected in our system. Interestingly, we excite the asymmetric mode (Fig. \ref{fig3}a and \ref{fig4}a) in our system that usually appears as the second subradiant mode in subwavelength systems \cite{bettles2016enhanced, asenjo2017exponential, Shahmoon2017Cooperative}. In addition, the mode has zero intensity on the diagonal, featuring a symmetry axis that is shifted by 45 degrees relative to the $k_x$ axis. We now explain how our unique system exhibits a higher-order spatial mode than the conventional subwavelength array systems. The SiV as a symmetry defect in diamond follows the crystal symmetry -- the (1,1,1) axis. This determines the polarization of emitted photons which is either aligned with the (1,1,1) or perpendicular to it. Therefore, all the possible emitted polarizations have an equal distribution on the $x$ and the $y$ axes. This fixes a preferred axis that is diagonal in the $xy$ plane, which supports modes that are asymmetric relative to the diagonal symmetry axis. We verify this claim by calculating the collective states for the (1,1,1) polarization and find the asymmetric subradiant modes (see Fig. \ref{fig4}d,e) observed in the experiment as one of three first subradiant collective states. To further understand why these specific asymmetric modes were excited we simulate the emission of a single dipole oriented in the (1,1,1) direction below the diamond-air interface by FDTD (Fourier-Domain-Time-Domain, Lumerical software) and find an anisotropic emission distribution in the far field (see Fig. \ref{fig4}b). This anisotropic emission is also revealed in our measurements for uncoupled (non-collective) emission (Fig. \ref{fig4}b) detected while not filtering the normal directed emission in k-space. We also find a phase flip around the symmetry axis favoring the asymmetric modes rather than symmetric ones. We find that the unique emission associated with the SiV dipole direction forces an asymmetric phase on the emission (Fig. \ref{fig4}c) -- favoring the asymmetric subradiant mode excitation (as opposed to atomic experiments in which the basic symmetrical mode was excited \cite{Rui2020Subradiant}).
Another interesting feature that can be explained by the dipole orientation and its proximity to the interface is the difference in intensity of the two subradiant mode lobes (See Fig. \ref{fig4}a) which is also supported by our FDTD simulation.   

We now describe our theoretical analysis for collective emission achieved despite inhomogeneous broadening. As mentioned before our specific array consists of superatoms in each site, which causes inhomogeneous broadening of the linewidth. The reason for the superatom implantation is to overcome the frequency missmatch between emitters in different sites, which was expected to suppress the collective emission. To design our SiV array we first calculate the collective resonances by employing the dyadic Green's function while accounting for the expected detuning between array's sites (see section 2 in Supplementary). From this analysis we extract the emitter density required per site for overcoming the collective effect suppression by inhomogeneity.  

In order to explain the origin of coherent emission in our specific experiment, we calculate the collective states of a single-site superatom comprised of detuned atoms. When accounting for tens of detuned atoms in a central site for frequencies distributed with $\Gamma_{eff}$ width around the bare emitter frequency, we find the condition for collective superradiance presented in\cite{masson2022universality} is met. We calculate the variance of the collective decay rates for such a superatom with four possible radii (ranging between 30 and 60 nm) and a range of detunings relative to the bare emitter frequency. The analysis  displayed in Fig. \ref{fig4}f provides an estimation of percentage of superatoms that meet the superradiance condition (collective decay rates variance above 1), as a function of their spectral width ($\Gamma_{eff}$). Specifically, for the $\Gamma_{eff} \sim$ 30\,GHz extracted from our measurements (average for the different SiV lines, see section section 2.1 in Supplementary) and the known bare-SiV linewidth we find that about 20 percent of prepared superatoms meet the superradiance condition. This explains how we ignite collective emission to begin with when exciting the arrays to higher energy levels (rather than on-resonance excitation).

\section{Methods}\label{sec11}

\subsection{Sample characteristics}
To SiV array is prepared by implanting ions through a pre-fabricated PMMA mask into the diamond. An array site is determined by the apertures' size in the mask (typically 50 nm). The depth of the SiV is about 60nm below the diamond surface enabling the near field emission to be collected by an objective into free space. We prepare array of SiVs with two distinct symmetries: honeycomb that has showed enhanced reflection in an atom array experiment \cite{bettles2016enhanced} and a square array \cite{Rui2020Subradiant}. For information regarding the fabrication process see Supplementary Information.

\subsection{Setup for probing collective emission}
In the experiment we excite with a green laser (532nm) and collect the emission by a scanning confocal and k-space microscope. Specifically, the confocal microscope configuration has been used to characterize the emission of the arrays. In the setup the excitation source passes through a dichroic mirror, and is focused onto the sample by an objective with an NA of 0.81. An array of lenses are used to map the emission in k-space plane onto a detector. We then spatially filter the emission with directionality normal to the array. Finally, a motorized mirror is used to map the k-space image over a free-space slit of a spectrometer (Princeton instruments, HRS-750) to resolve the directional emission intensity profile and spectrum. 

\subsection{Theoretical methods}
The collective resonance of the atom-like array is solved by diagonalizing numerically the effective Hamiltonian. The system of an atom array with sub-wavelength spacing can be described by an effective Hamiltonian after tracing out the optical degrees of freedom in the Born--Markov approximation 
:
\begin{equation}
\begin{aligned}
H_{\mathrm{eff}} = \sum^{N}_{i=1}{\omega_{i}\hat{\sigma}^{\dagger}_{i}\hat{\sigma}_{i}}
+ \sum_{i,j}{(J_{ij} - i\frac{\Gamma_{ij}}{2}) \cdot \hat{\sigma}_{i}^{\dagger}\cdot \hat{\sigma}_{j}}\,,
\label{eq:3}
\end{aligned}
\end{equation}

where $ij$ runs over the atoms, $\omega_{i}$ is the bare resonance frequency, and $G_{ij}$ the dyadic Green function that captures the exchange of radiation between the systems \cite{Novotny2012}, $\Gamma_{ij}$ is the lifetime and $J_{ij}$ the resonance frequency. 
The eigenstates obtained from the diagonalization of the Hamiltonian represent the excitation probability of a specific atom in the array, allowing us to characterize the spatial distribution of the collective modes (such as the one presented in Figs. \ref{fig4}d,e). To simulate our specific system we consider an ensemble of 10s of emitters, typical distances of around 30-60\,nm and frequencies gaussian distributed with standard deviation of $\sim 10$ Ghz around the bare emitter resonance frequency. The positions of the emitters and their frequencies were randomly sampled in each run from the distributions specified above, and the probability of obtaining a collective superradiant effect was calculated as the number of runs that meet the collective emission condition (out of 1000 runs for each value of our superatom linewidth $\Gamma_{eff}$).


\section{Conclusion}\label{sec13}

To conclude, we have realized a metamaterial composed of quantum emitters that emit light coherently, with a spectrum and specific directionality that is determined by the pre-designed array symmetry and emitter properties. We have done so with solid-state emitters in which inhomogeneity was originally expected to suppress any collective effects. 

Our work opens the door for bottom-up quantum metamaterials in the scalable SiV-centre platform \cite{ding2024high}, by demonstrating a method for solid-state-based metamaterials overcoming the basic limiting factor --- the inhomogeneity.
The method presented can be adapted to any other quantum emitters that suffer limitations due to inhomogeneity. This novel capability, combined with the quantum properties and control of silicon-vacancy centres, will enable a scalable platform for the generation of coherence between solid-state emitters and photons. For example, employing quantum control of the SiV's electronic and nuclear spins \cite{Bhaskar2020experimental,stas2022robust} for generation of many-body photonic state \cite{Pichler2017universal,borregaard2020one}, or generation of quantum light that can be integrated into nanophotonics \cite{knall2022efficient} due to the coherent emission directionality reported.

\backmatter


\bmhead{Acknowledgements}
This research was funded by Grant No. 2021775 from the United States-Israel Binational Science Foundation (BSF) and by Grant No. 2207972. From the United States National Science Foundation (NSF) and ISF Grant No. 2402/22. This work has received funding from the European Research Council (ERC) under the European Union's Horizon 2020 research and innovation programme (Grant Agreement No. 101117845).

\section*{Declarations}

\textbf{Competing Interests Statement} The authors declare no competing interests.\\
\textbf{Author contribution} U.I, S.L and S.B have contributed equally to this work. U.I, S.L and S.B. A.K, D.N and R.B have contributed to all aspects of the work. S.W has contributed to the theoretical analysis. \\
\textbf{Data availability} Additional data, as well as details of the numerical and experimental methods are available upon request.






\bibliography{sn-bibliography}

\end{document}